\documentclass[
twocolumn,times,
twocolappendix
]{aastex701}
\usepackage{apjfonts}

\usepackage{xspace}
\usepackage{xcolor}
\newcommand{\event}{GW200105\xspace}
\newcommand{\eGR}{e_{\rm GR}\xspace}
\newcommand{\eObs}{e_{\rm obs}\xspace}

\newcommand{\efpe}{\texttt{pyEFPE}\xspace}

\begin{document}

\title{Testing the nature of GW200105 by probing the frequency evolution of eccentricity}

\newcommand{\IUCAA}{\affiliation{Inter-University Centre for Astronomy and
    Astrophysics, Post Bag 4, Ganeshkhind, Pune 411007, India}}
\newcommand{\SNU}{\affiliation{Department of Physics and Astronomy,
    Seoul National University, Seoul 08826, Korea}}
\newcommand{\VSM}{\affiliation{Department of Physics, Vivekananda Satavarshiki Mahavidyalaya (affiliated to Vidyasagar University),\\ Manikpara 721513, West Bengal, India}}

\author[0000-0001-7197-8899]{Avinash Tiwari}
\email{avinash.tiwari@iucaa.in}
\IUCAA
\author[0000-0002-6783-1840]{Sajad A. Bhat}
\email{sajad.bhat@iucaa.in}
\IUCAA
\author[0000-0003-0826-6164]{Md Arif Shaikh}
\email{arifshaikh.astro@gmail.com}
\VSM
\author[0000-0001-5318-1253]{Shasvath J. Kapadia}
\email[show]{avinash.tiwari@iucaa.in, sajad.bhat@iucaa.in,\\
arifshaikh.astro@gmail.com, shasvath.kapadia@iucaa.in}
\IUCAA

\begin{abstract}
GW200105 is a compact binary coalescence (CBC) event, consisting of a neutron
star and a black hole, observed in LIGO–Virgo–KAGRA's (LVK's) third observing
run (O3). Recent reanalyses of the event using state-of-the-art waveform models have claimed observation of signatures of an eccentric orbit. It has nevertheless been pointed out in the literature that certain physical or modified gravity effects could mimic eccentricity by producing a spurious non-zero eccentricity value, at a given reference frequency, when recovered with an eccentric waveform model. We recently developed a waveform-model-independent Eccentricity Evolution Consistency Test \citep[EECT,][]{Bhat:2025lri} to identify such mimickers, by comparing the measured frequency {\itshape evolution} of eccentricity, $e(f)$, with that expected from General Relativity (GR). In this {\itshape paper}, we apply EECT to GW200105 and find that it satisfies EECT within 68\% confidence. Our analysis therefore lends complementary support in favour of the eccentricity hypothesis, while also providing a novel test of the consistency of $e(f)$ with GR.
\end{abstract}



\section{Introduction}\label{sec:introduction}
\event \citep{2021ApJ...915L...5A} is a neutron star black hole (NSBH) merger event observed by LIGO-Livingston and Virgo. It is one of the 218 compact binary
coalescences (CBCs) detected till the first part of the fourth observing run \citep[O4a,][]{LIGOScientific:2025slb} of the LIGO-Virgo-KAGRA (LVK) network of gravitational wave (GW)
detectors \citep{TheLIGOScientific:2014jea, TheVirgo:2014hva, KAGRA:2020tym}. The standard expectation is that the orbits of CBCs are circularized due to loss of energy and angular momentum via
GW emission~\citep{Peters:1963ux}. Consequently, any residual eccentricities of the GWs when they enter the frequency band of the detectors are too small to be observable. This argument, in tandem with the lack of reliable eccentric waveform templates, justified the use of quasi-circular templates for detection and parameter estimation (PE) of events in the LVK's third Gravitational-Wave-Transient-Catalog \citep[GWTC-3,][]{KAGRA:2021vkt}. On the other hand, certain dense stellar environments could harbour eccentric CBCs \citep{Mapelli:2021for}, due to mechanisms such as Kozai-Lidov excitations \citep{Kozai:1962zz, Lidov:1962wjn, Naoz:2016tri}. In fact, even hierarchical triples in the galactic field could induce mergers with measurable eccentricities, in future and (possibly) current detectors as well \citep[see, e.g.,][]{Dorozsmai:2025jlu}. Thus, the identification of signatures of eccentricity in detected CBCs is of particular interest to understand and constrain CBC formation channels.

In recent years, several eccentric waveform models have become available \citep{Gamboa:2024imd,Gamboa:2024hli,Nagar:2024dzj, Gamba:2024cvy, Albanesi:2025txj,Paul:2024ujx, Planas:2025feq,Morras:2025nlp, Islam:2021mha, Islam:2024rhm, PhysRevD.111.L081503, Islam:2024zqo, pk8n-fxvw, Islam:2025llx}. These advances in eccentric waveform construction and generation have enabled large-scale PE on GWTC-3 events to search for signatures of eccentricity. Independent analyses using some of the waveform models have identified multiple events exhibiting signatures of nonzero eccentricity at a given reference frequency \citep{Gupte:2024jfe, Planas:2025jny, Morras2025_GW200105_ecc,
Planas:2025plq, Romero-Shaw:2021ual, Romero-Shaw_2022}. In particular, the recent identification of \event as an eccentric event by \cite{Morras2025_GW200105_ecc} using the eccentric spin-precessing Post-Newtonian waveform model \efpe \citep{Morras:2025nlp} has reignited significant interest in \event. Subsequent reanalyses of \event using different waveform models have provided further support for the eccentric nature of \event \citep{Planas:2025plq,Wang:2025yac,Kacanja:2025kpr,Roy:2025xih,Jan:2025fps}.

On the other hand, a number of physical or beyond-GR effects could either imitate or be mimicked by eccentricity. These include (but are not limited to) microlensing of GWs exhibiting wave-optics effects \citep{mishra-ecc-mclz}, line-of-sight acceleration \citep{Tiwari:2025aec}, spin-precession \citep{Romero-Shaw:2022fbf}, massive graviton effect \citep{Will:1997bb, Narayan:2023vhm}, and dipole radiation \citep{Will:2014kxa, PhysRevLett.116.241104, PhysRevD.86.022004, Narayan:2023vhm}. This necessitates a method to identify truly eccentric CBCs from quasi-circular ones with modulations driven by physics unrelated to eccentricity. The conventional approach adopts Bayesian model selection, requiring large-scale PE runs that sample the GW PE posterior under different hypotheses, including the eccentricity hypothesis. Such an approach is not only computationally expensive and time-consuming, but could also be potentially misleading if none of the hypotheses considered correspond to the underlying physics of the GW signal. Moreover, a number of hypotheses may not have readily available waveform models that can be used for GW PE, making Bayesian model selection either computationally unfeasible or outright (currently) impossible. 

In our previous work \citep{Bhat:2025lri}, we proposed an Eccentricity
Evolution Consistency Test (EECT) to distinguish between genuinely eccentric
signals from those that mimic them. The method rests on the following
expectation -- while eccentricity mimickers can produce a non-zero eccentricity
at some given reference frequency when recovered with an eccentric waveform
model, they may not, in general, imitate its evolution with frequency. EECT
accordingly compares eccentricities recovered at certain frequencies to those
at the same frequencies expected from the GR-predicted frequency evolution of
eccentricity. We demonstrated the power of EECT by applying it to various
eccentricity mimickers that produce spurious non-zero eccentricities at given
reference frequencies, but fail the test (at $> 68\%$ confidence) due to their
inability to mimic the GR-prescribed frequency evolution. On the other hand,
EECT applied to truly eccentric signals exhibits no such violation (i.e, they
satisfy EECT within $68\%$ confidence).

In this work, we apply, for the first time, EECT to \event, as well as a \event-like zero-noise injection. We find that in both cases, EECT is satisfied within $68\%$ confidence. This lends complementary support in favor of the eccentricity hypothesis for this event, and demonstrates \event's consistency with the GR-predicted evolution of eccentricity. 

\section{Summary of Method}\label{sec:method}
 Let $\eObs (f)$ be the eccentricity recovered at some GW frequency $f$ and
 $\eGR (f)$ be the GR-predicted eccentricity acquired by evolving $\eObs(f =
 f_0)$ from some initial frequency $f_0$. Then the eccentricity deviation
 parameter $\delta_e (f)$ is define as \citep{Bhat:2025lri}:
\begin{equation}
    \label{eq: dev_par}
    \delta_e(f) \equiv 2 \frac{\eGR (f) - \eObs (f)}{\eGR (f) + \eObs (f)} 
\end{equation}

To construct the posteriors of $\delta_e (f)$, we adopt the following prescription:
\begin{itemize}
    \item Recover the eccentricities $\eObs (f)$ at some reference frequencies,
      say $f = \lbrace { f_{\rm ref} \rbrace}$, by keeping the minimum ($f_{\rm
        min}$) and reference frequencies ($f_{\rm ref}$) equal to each other. Note that
      $f_{\rm min}$ is the lower limit of the  frequency range over which the
      GW likelihood is evaluated. Changing $f_{\rm min}$ is crucial, as
      explained in \citet{Bhat:2025lri}. In this  work, we set the upper limit
      $f_{\rm max}$ to 1792 Hz.
    
    \item Ensure $\eObs (f)$ posteriors are sufficiently deviated from zero --
      i.e, zero is excluded at some predefined confidence level. In this {\itshape
        paper}, we choose this threshold to be 90\%.
    \item Evolve $\eObs (f_0)$ to $\eGR (f_{\rm ref})$ assuming GR. Note that
      different GR-consistent waveform models may differ from each other
      slightly. To avoid waveform-driven systematics, use the model-prescribed
      frequency evolution of eccentricity.~\footnote{Different waveform models
        may also differ in their eccentricity
        definitions~\citep{Shaikh:2023ypz,Shaikh:2025tae, islam2025pntifc}. However, as long as the same waveform model is used for $\eObs$ and $\eGR$, EECT should hold
      irrespective of the choice of eccentricity definition within the model.}
    \item To ensure that $\eObs (f)$ and $\eGR (f)$ have equivalent priors, reweight the evolved posterior on $\eGR (f)$ to the prior used to construct the posteriors on $\eObs (f)$.
    \item Use Eq.~\ref{eq: dev_par} to construct the posteriors of $
    \delta_e (f)$ at different $f = \left\{f_{\rm ref}\right\}$.
\end{itemize}
Once constructed, a truly eccentric signal is expected to have the posteriors
on $\delta_e(f)$ consistent with $0$ within a confidence interval. In this
work, we set the width of this interval to $90\%$. If $\delta_e(f)$ is deviated from zero at $> 90\%$ confidence, EECT is violated, suggesting the presence of an eccentricity mimicker, or possibly a violation of GR. We refer the reader to our methods paper \citep{Bhat:2025lri} for a detailed exposition of the method and its application to identifying mimickers and testing GR.

\section{Results}\label{sec:results}
We present results of applying EECT to \event, following the prescription described in the previous section, and using the \efpe \citep{Morras:2025nlp} waveform. \efpe models the inspiral and incorporates eccentricity and spin-precession effects together with the higher eccentric harmonics. However, it does not include the merger-ringdown phase, nor does it model NS tidal effects. Nevertheless, \efpe has been argued to be applicable to \event, because the signal is dominated by the inspiral with no appreciable signatures of the merger-ringdown phase. Effects of NS tides on the signal are also expected to be suppressed \citep{2021ApJ...915L...5A}. See \citet{Morras2025_GW200105_ecc} for additional details on the applicability of \efpe for PE on \event. 

\begin{figure*}[htb]
    \centering
    \includegraphics[width=0.495\linewidth]{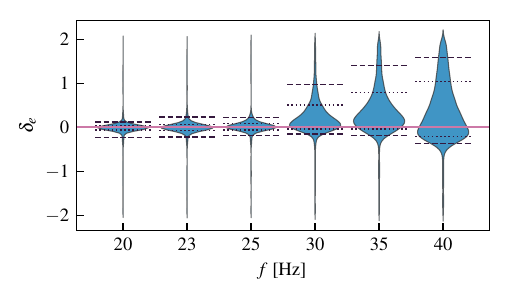}
    \includegraphics[width=0.495\linewidth]{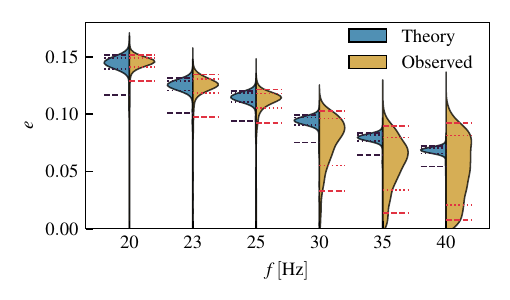}
    \caption{\textit{Left:} Violins of the eccentricity deviation $\delta_e$ at different reference frequencies. The horizontal line at $\delta_e = 0$ represents zero deviation from GR. \textit{Right:} Individual $\eObs$ (right side) and $\eGR$ (left side) posteriors at different reference frequencies for the same. In both panels, the dashed and the dotted lines represent the 90\% and 68\% credible intervals (CI), respectively.}
    \label{fig: violin_plot_ps}
\end{figure*}

We sample the 17-dimensional GW posterior using detector data surrounding \event, acquired from the Gravitational Wave Open Science Center (GWOSC)~\citep{KAGRA:2023pio}. The parameters sampled, as well as the corresponding priors used, are tabulated in Table~\ref{tab:priors-real} of Appendix~\ref{sec:app1}. We do so for multiple minimum and reference frequencies:~\footnote{We find that our PE results for $f_{\rm min} = f_{\rm ref} = 20$ Hz are fully consistent with those of \citet{Morras2025_GW200105_ecc}, as expected, given that we use the same waveform model.} $\lbrace{f_{\rm min}\rbrace} = \lbrace{f_{\rm ref}\rbrace} = \lbrace{18, 20, 23, 25, 30, 35, 40 \rbrace}$ Hz~\footnote{The frequencies are chosen such that the median value of the eccentricity posterior at $f_0$ drops by, roughly, $0.02$ at each $f_{\rm ref}$ when evolved according to GR following \citet{Peters:1963ux}.}. We then select $f_0 = 18$ Hz, and evolve $\eObs(f_0)$ to $\eGR(\lbrace{f_{\rm ref}\rbrace})$. The evolution assumes that eccentricity decays under GW radiation alone as predicted by GR, with no other imprints of physical or beyond-GR effects. We ensure that, for the reference frequencies chosen, $\eObs$ posteriors exclude zero at $> 90\%$ confidence, and all $\eGR$ posteriors are reweighted as described in the previous section.

From the $\eObs$ and $\eGR$ posteriors, we construct posteriors on $\delta_e$ at the same set of reference frequencies as above. These are presented as violins in the \textit{left} panel of Figure~\ref{fig: violin_plot_ps} at each of the reference frequencies chosen, except $f_{\rm ref} = f_0 = 18$ Hz where $\delta_e$ is consistent with zero by design. A complementary \textit{right} panel in Figure~\ref{fig: violin_plot_ps}  shows the individual $\eObs$ and $\eGR$ posteriors. Both figures demarcate the $68\%$ (dotted) and $90\%$ (dashed) confidence interval for all the posteriors displayed.

The $\delta_e$ posteriors in the \textit{left} panel of Figure~\ref{fig: violin_plot_ps} are found to be consistent with zero within $90\%$ confidence at all chosen reference frequencies. Indeed, they're also found to be consistent within $68\%$, with no compelling evidence of violation of EECT. The \textit{right} panel of Figure~\ref{fig: violin_plot_ps} further corroborates this visually by showing the consistency between the individual $\eObs$ and $\eGR$ posteriors.

\begin{figure*}
    \centering
    \includegraphics[width=0.495\linewidth]{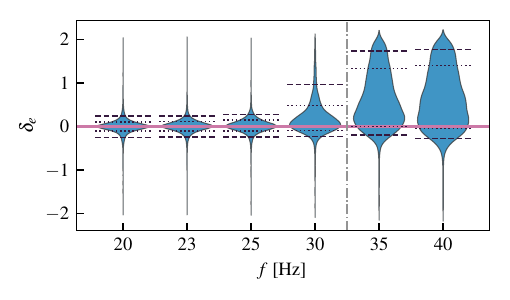}
    \includegraphics[width=0.495\linewidth]{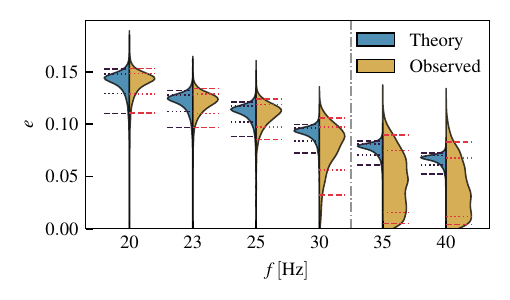}
    \caption{Same as Figure~\ref{fig: violin_plot_ps}, but for a \event-like injection. {\itshape Left:} Violins of eccentricity deviation $\delta_e$ at different reference frequencies. The horizontal line at $\delta_e = 0$ represents zero deviation from GR. \textit{Right}: Individual $\eObs$ (right side) and $\eGR$ (left side) posteriors at different reference frequencies for the same. In both panels, the dashed and the dotted lines represent the 90\% and 68\% credible intervals (CI), respectively. The vertical dash-dotted gray lines between 30 and 35 Hz indicate the onset of $\eObs$ posteriors (35 and 40 Hz ones) that fail the zero-exclusion criterion described in Section~\ref{sec:method}.}
    \label{fig: violin_plot_as_inj}
\end{figure*}

In Figure~\ref{fig: violin_plot_as_inj}, we present posteriors on $\delta_e$ (\textit{left} panel) for a \event-like zero-noise injection with injection parameters: chirp mass $\mathcal{M} = 3.538\,  M_{\odot}$, mass ratio $q = 0.132$, spin-magnitudes $a_1 = 0.055$ and $a_2 = 0.239$, tilts $\theta_1 = 1.618$ and $\theta_2 = 1.569$, spin–spin azimuthal angle $\phi_{12} = 3.172$, precession phase angle $\phi_{JL} = 3.161$, declination $\delta = -0.021$, right ascension ${\rm RA} = 3.959$, viewing angle $\theta_{JN} = 2.6$, polarization angle $\psi = 1.578$, geocentric time $t_{\rm c} = 1262276684.057 \, \rm s$, luminosity distance $d_{\rm L} = 306.443 \, \rm Mpc$, phase at the reference frequency $\phi = 3.176$, eccentricity $e = 0.161$, mean anomaly $\ell = 3.018$, assuming a 2-detector (L1, V1)~\footnote{LIGO-Livingston, Virgo} network and O3-like noise PSDs. We have used the same priors as the real event, which are tabulated in Table~\ref{tab:priors-real} of Appendix~\ref{sec:app1}.

A complementary \textit{right} panel in Figure~\ref{fig: violin_plot_as_inj} shows the individual posteriors on $\eObs$ and $\eGR$. As with EECT applied to \event, when applied to this \event-like injection, the test is satisfied not only within $90\%$ confidence, but also within $68\%$. This acts as an important confirmation that a truly eccentric \event would satisfy EECT. Note that, even though the $\delta_e$ violins at $f_{\rm ref} = 35$ Hz and 40 Hz contain zero within $68\%$ confidence, the eccentricity posteriors $\eObs$ at these frequencies violate the zero-exclusion criterion described in Section~\ref{sec:method}. Drawing any conclusions regarding EECT for this injection, at these frequencies, should be avoided. We have put a vertical dash-dotted gray line between 30 and 35 Hz to demarcate the same. 

To ensure that the consistency of \event with GR-predicted eccentricity evolution is not a result of the well-known correlation between chirp mass and eccentricity, we also plot the recovered detector-frame chirp mass posteriors at $\lbrace{f_{\rm ref}\rbrace}$ in Figure~\ref{fig: mc_post} of Appendix~\ref{sec:app2}. We find that these are all consistent with each other, as expected. For completeness, we also present a corner plot of the posteriors on all the \event parameters, at two reference frequencies, in Figure~\ref{fig: corn_plot_comb} of Appendix~\ref{sec:app2}. All posteriors of parameters that are not expected to evolve with frequency are found to be consistent with each other.

\section{Discussion}

We applied EECT -- developed and demonstrated in \cite{Bhat:2025lri} -- to \event, and found that it is satisfied within $90\%$ (and $68\%$) confidence, at all reference frequencies considered. A \event-like zero-noise injection assuming a 2-detector-network (L1, V1) operating at O3-like sensitivity, was also found to satisfy EECT within the same confidence intervals. Our work, therefore, adds complementary support in favor of the eccentricity hypothesis for this event, with no evidence to suggest the presence of an eccentricity mimicker. It is the first waveform-model-independent test of the eccentric nature of \event's orbit, while also demonstrating consistency of the signal with GR-predicted frequency evolution of eccentricity, thus acting as a novel test of GR on this event. 

The advantage of EECT, over conventional Bayesian model-selection methods, is amply demonstrated here. First, we did not require the plethora of alternative hypotheses to ascertain that the eccentricity hypothesis is favored over others, thus saving on time and computational costs. Moreover, the uncertainty of whether {\it all} possible hypotheses that could mimick eccentricity have been considered, does not apply here, as it would in a Bayesian model selection approach.  

It should be pointed out that there could be physical and beyond-GR effects that are too subtle to be captured as deviations from the GR-expected eccentricity evolution, in the LVK detector network's O4. Moreover, EECT becomes less sensitive to deviations with increasing reference frequencies, where posteriors on $\eObs$ broaden due to reduced SNR and number of in-band cycles -- as is observed in Figure~\ref{fig: violin_plot_ps}. Thus, it is conceivable that violations of EECT that should have manifested at frequencies $> 40$ Hz, are missed because those are rejected by virtue of $\eObs$ not satisfying the zero-exclusion criterion (cf. Section~\ref{sec:method}). Nevertheless, with improved detector range and bandwidth, as is expected in future observing scenarios, such deviations will also become accessible to EECT. 

We draw the reader's attention to a point of contention regarding the choice of prior on $\eObs$. Some works on probing the eccentric nature of \event have shown that choosing a log-uniform prior on eccentricity, instead of a uniform one like we do in this work, leads to a reduced Bayes factor in favor of the eccentricity hypothesis versus the quasi-circular one. Indeed, even the value of eccentricity recovered at $20$ Hz reduces when a log-uniform prior is used \citep{Planas:2025plq, Planas:2025jny, Kacanja:2025kpr, Jan:2025fps}. Nevertheless, as has been argued in \citet{Morras2025_GW200105_ecc}, a log-uniform prior is one that is potentially over-informed. It favors lower eccentricities and thus biases their estimates. 

We end by recommending EECT as a crucial probe of the true nature of any event that seems to exhibit signatures of eccentricity at a given reference frequency. Any claim regarding the eccentric nature of an event must be supported by the results of EECT to ascertain that a mimicker is not manifestly at play. We also encourage its use when probing the GR nature of eccentricity evolution. Furthermore, large-scale Bayesian model-selection enterprises should only be embarked upon once EECT has been applied and the corresponding results considered.

\begin{acknowledgements}
  The authors would like to thank Isobel Romero-Shaw for valuable comments on the draft. M. A. S. acknowledges hospitality at IUCAA during his visit related to this work. S.J.K. acknowledges support from ANRF/SERB grants SRG/2023/000419 and MTR/2023/000086. 
  
  This research has made use of data or software obtained from the Gravitational Wave Open Science Center (gwosc.org), a service of the LIGO Scientific Collaboration, the Virgo Collaboration, and KAGRA. This material is based upon work supported by NSF's LIGO Laboratory which is a major facility fully funded by the National Science Foundation, as well as the Science and Technology Facilities Council (STFC) of the United Kingdom, the Max-Planck-Society (MPS), and the State of Niedersachsen/Germany for support of the construction of Advanced LIGO and construction and operation of the GEO600 detector. Additional support for Advanced LIGO was provided by the Australian Research Council. Virgo is funded, through the European Gravitational Observatory (EGO), by the French Centre National de Recherche Scientifique (CNRS), the Italian Istituto Nazionale di Fisica Nucleare (INFN) and the Dutch Nikhef, with contributions by institutions from Belgium, Germany, Greece, Hungary, Ireland, Japan, Monaco, Poland, Portugal, Spain. KAGRA is supported by Ministry of Education, Culture, Sports, Science and Technology (MEXT), Japan Society for the Promotion of Science (JSPS) in Japan; National Research Foundation (NRF) and Ministry of Science and ICT (MSIT) in Korea; Academia Sinica (AS) and National Science and Technology Council (NSTC) in Taiwan.

\end{acknowledgements}

\bibliography{references}{}
\bibliographystyle{aasjournalv7}

\appendix

\section{Priors}\label{sec:app1}
Priors for PE of \event are listed in 
Table~\ref{tab:priors-real}. 

\begin{deluxetable}{ll}
    \caption{Priors used for PE of \event. $\mathcal{U}(a, b)$ refers to uniform distribution in the range $(a, b)$ while PL$(a,b)$ refers to a power law distribution in the range $(a,b)$.}
    \label{tab:priors-real}
\tablehead{
\colhead{Parameter} & \colhead{Prior}
}
\startdata
        Chirp mass $\mathcal{M} \, [\rm M_{\odot}]$   & $\mathcal{U}(3.2, \, 4.0)$ \\
        Mass ratio $q$   & $\mathcal{U}$(0.05, 0.5)  \\
        Eccentricity $e$   & $\mathcal{U}$(0, 0.4)  \\
        Mean Anomaly $\ell$ [rad]  & $\mathcal{U}$(0, $2\pi$)  \\
        Luminosity distance $d_{\rm L} \, [{\rm Mpc}]$  & PL(10, 2000) $\propto d_{\rm L}^2$ \\
        Dimensionless spins $a_{1,2}$ & $\mathcal{U}(0, 0.5)$ \\
        Tilts $\theta_{1,2}$ & $\sin$(0, $\pi$) \\
        Spin–spin azimuthal angle $\phi_{12}$ [rad] & $\mathcal{U}$(0, $2\pi$) \\
        Precession phase angle $\phi_{\rm JL}$ [rad] & $\mathcal{U}$(0, $2\pi$) \\
        Right Ascension [rad] & $\mathcal{U}$(0, $2\pi$) \\
        Declination $\delta$ & $\sin$(0, $\pi$) \\
        Viewing angle $\theta_{\rm JN}$ & $\sin$(0, $\pi$) \\
        Polarization angle $\psi$ [rad] & $\mathcal{U}$(0, $\pi$) \\
        Phase at the reference frequency $\phi$ [rad] & $\mathcal{U}$(0, $2\pi$) \\
        Geocent time $t_{\rm c}$ [s] & $\mathcal{U}(1262276683.957,$ \\
        & $~~~~~1262276684.157)$
        \enddata
\end{deluxetable}

\section{Additional Figures}\label{sec:app2}
\begin{figure}
    \centering
    \includegraphics{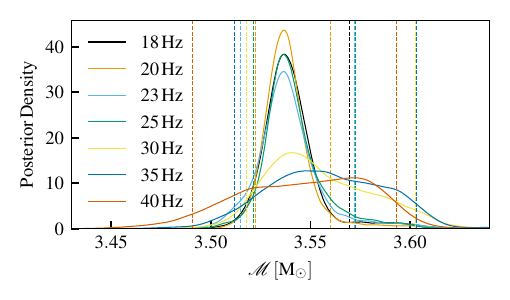}
    \caption{Detector-frame chirp mass posteriors recovered at different reference frequencies.}
    \label{fig: mc_post}
\end{figure}

\begin{figure*}
    \centering
    \includegraphics[width=\textwidth]{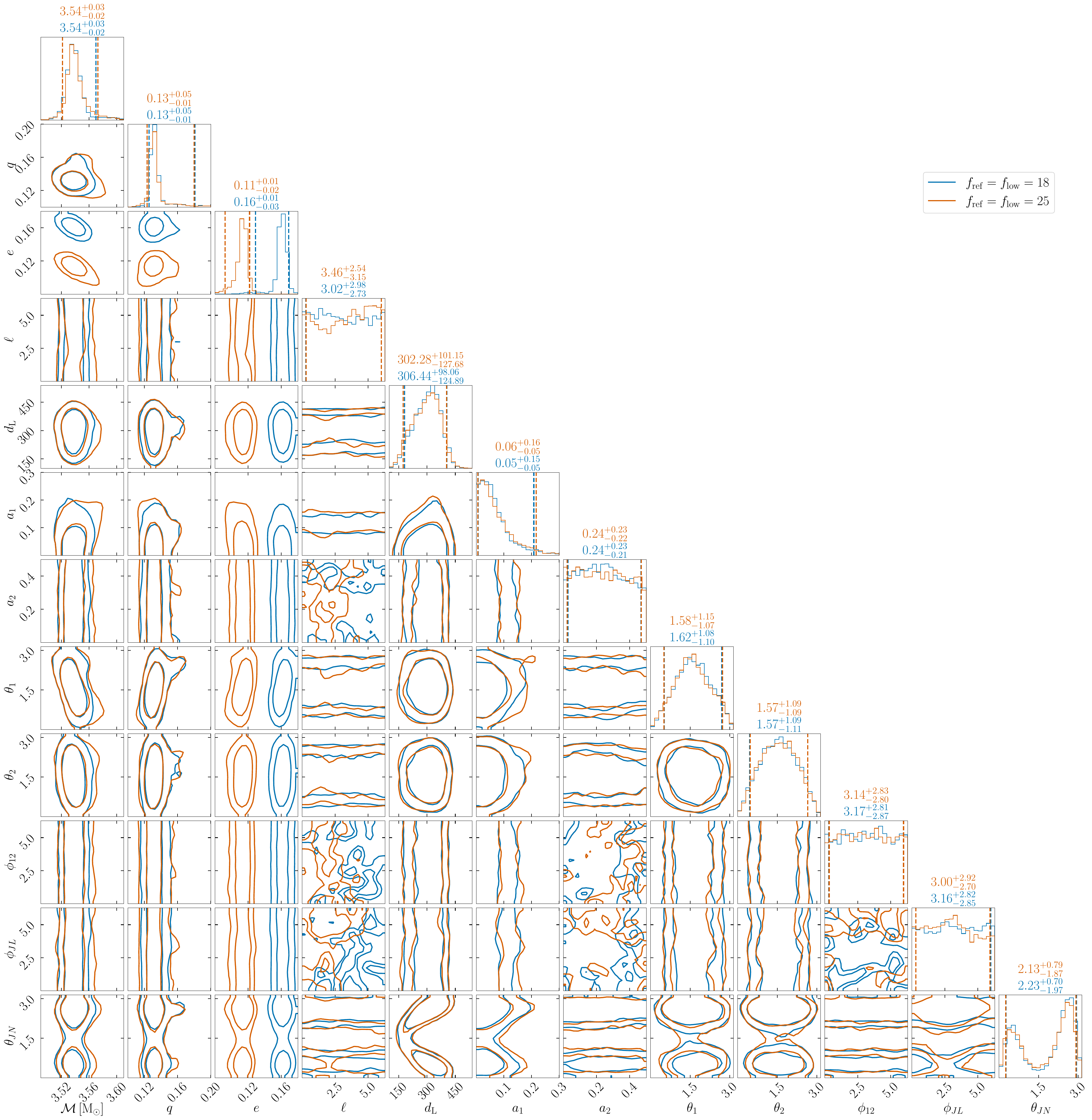}
    \caption{Corner plot of posteriors recovered at 18 Hz (blue) and 25 Hz (maroon) for \event.}
    \label{fig: corn_plot_comb}
\end{figure*}

To show that the recovered chirp mass is not changing with reference frequencies, in Figure~\ref{fig: mc_post}, we show the detector-frame chirp mass posteriors of \event recovered at different reference frequencies $\lbrace{f_{\rm ref} \rbrace}$. Additionally, for completeness, we also show posteriors on all source parameters --  except Right Ascension, Declination, $\psi$, $\phi$, and $t_{\rm c}$ -- of \event, recovered at two different frequencies, viz. 18 Hz and 25 Hz, in Figure~\ref{fig: corn_plot_comb}. 

\end{document}